\documentclass[letterpaper]{article}
\usepackage{aaai}
\usepackage{times}
\usepackage{helvet}
\usepackage{courier}
\usepackage{comment}
\usepackage{bbm}
\usepackage{diagbox}
\usepackage{adjustbox}
\usepackage{newtxtext,newtxmath}
\usepackage{url}
 \usepackage{soul}
\usepackage{graphicx}
\graphicspath{ {./Images/} }
\usepackage{caption}
\usepackage{subcaption}
\usepackage{amsmath}
\usepackage{mathtools, nccmath}
\usepackage{tikz}
\usetikzlibrary{shapes,arrows, fit,calc}
\usetikzlibrary{shadows}
\usetikzlibrary{patterns}
\usetikzlibrary{positioning}
\usepackage{xcolor,colortbl}

\definecolor{LightCyan}{rgb}{0.88,1,1}

\newcolumntype{a}{>{\columncolor{LightCyan}}c}

\begin{document}
%
\title{Can we infer player behavior tendencies from a player's decision-making data? Integrating Theory of Mind to Player Modeling}
\author{
Murtuza N. Shergadwala,\textsuperscript{\rm 1}
Zhaoqing Teng,\textsuperscript{\rm 1}
Magy Seif El-Nasr,\textsuperscript{\rm 1}\\

\textsuperscript{\rm 1}University of California, Santa Cruz\\
mshergad@ucsc.edu, zhteng@ucsc.edu, mseifeln@ucsc.edu
}






\maketitle
\begin{abstract}
\begin{quote}
Game AI systems need the theory of mind, which is the humanistic ability to infer others' mental models, preferences, and intent. Such systems would enable inferring players' behavior tendencies that contribute to the variations in their decision-making behaviors. To that end, in this paper, we propose the use of inverse Bayesian inference to infer behavior tendencies given a descriptive cognitive model of a player's decision making. The model embeds behavior tendencies as weight parameters in a player's decision-making. Inferences on such parameters provide intuitive interpretations about a player's cognition while making in-game decisions. We illustrate the use of inverse Bayesian inference with synthetically generated data in a game called \textit{BoomTown} developed by Gallup. We use the proposed model to infer a player's behavior tendencies for moving decisions on a game map. Our results indicate that our model is able to infer these parameters towards uncovering not only a player's decision making but also their behavior tendencies for making such decisions.

\end{quote}
\end{abstract}

\section{Introduction}


In this study, we consider the problem of computationally identifying behavior tendencies from a player's game decision-making data. By ``behavior tendencies'' we refer to a player's consideration of the relevance of attributes that helps them make a specific decision. For example, a player's decision to move on a map can be influenced by several situational factors or attributes such as the location of a valuable resource or the presence of a threat. Each player has their own subjective characteristics that determine the relevance of such attributes that ultimately influence the variations in their decision making behaviors across a player population. 
 
Humans can easily identify such behavior tendencies by observing someone's gameplay -- an ability called the Theory of Mind~\cite{premack1978does}, which is often attributed to successful collaboration in teams \cite{engel2014reading} and other environments. While such capability is important, game AI agents/characters are not developed with it. While there has been much work on developing algorithms to infer plans, goals, or personality from gamelog data e.g., \cite{bunian2017modeling,baker2009action,albrecht1998bayesian}, to mention a few, much of this work face various challenges. First, inferring behavior tendencies, intent or preferences require inferring latent (cognitive) variables that are not observed through game logs or game data, which requires probabilistic modeling. Second, such variables are player specific which makes it further difficult to generalize inferences across a player population. This makes it hard to use off-the-shelf machine learning techniques without further modeling. Third, there is a lack of theory driven player models that are required to make inferences on latent variables in an explainable and meaningful manner. While cognitive science have made various strides, the complexity of game environments accompanied by the need to integrate many different cognitive processes to explain players' problem-solving process makes it hard to apply current cognitive models without an integrative approach.   

In this paper we address this gap by specifically targeting the research question of: \textit{How can we infer player-specific tendencies that influence their decision-making behaviors in a digital game?} To address this question, we develop a simple yet explainable probabilistic player model to simulate a player's 2-sequence decision $(\textit{D1}, \textit{D2})$, where a player makes decision $\textit{D1}$ then as a consequence will need to make a decision $\textit{D2}$. We limited the decision model to two consecutive decisions as a starting point, which we aim to expand in future work. We then leverage inverse Bayesian inference to infer model parameters given synthetically generated data for agents with varying behavior tendencies. We verify the inferred parameters with the actual parameter values. 

Our contribution includes the proposed model and an illustration of the model contextualization and implementation via a use case game called \textit{BoomTown} developed by Gallup. The objective of the game is simple: maximize the amount of gold (resource) collected through mining in a given map with rocks and gold. In such a scenario, we consider two behavior tendencies: (1) rock agnostic tendency, and (2) a rock aversion tendency. A player with rock agnostic tendency would attribute much consideration to large gold clusters and would not care about the amount of rock structures surrounding the gold clusters. On the other hand, a player with rock aversion tendency focuses more on the gold in the rock-free regions such that they would not have to mine through the rocky mountains to reach to the gold. Our approach enables us to model such behavior tendencies which can be extended to other games and gameplay contexts.

\section{Related Work} \label{sec:lit_rev}

Existing work on player modeling can be categorized as ``generative'' or ``descriptive'' based on the purpose of modeling~\cite{smith2011inclusive}. Generative purpose of player modeling focuses more on producing simulations of human player~\cite{yannakakis2013player,machado2011player,van2005opponent,machado2011player,machado2011player2}. Whereas, descriptive purpose of player modeling focuses more on describing player's decisions, behaviors, and preferences~\cite{andersen2010gameplay,holmgaard2015monte,holmgaard2014personas,bindewald2016clustering,cowling2014player}. In this paper, we focus on descriptive player modeling. However, we note that the vocabulary provided by Smith et. al.~\cite{smith2011inclusive} to classify descriptive player models insinuates that descriptive models are intended to provide ``high-level description" of player behavior. On the contrary, we illustrate modeling the decision-making process of the players at a granular level such that it abstracts a player's cognitive processes while making game-specific decisions. Thus, there is a lack of descriptive and computational cognitive models of player behaviors that are explainable towards understanding a player’s behavioral tendencies. Consequently, we focus on related work in plan recognition within games that closely matches the categorization of descriptive modeling with granular details such as a player's decision making process.

Player plan recognition refers to algorithms that focus on computationally recognizing a player's behaviors, strategies, goals, plans, and intent. Work in this area is summarized in~\cite{el2016game,el2021game}. The approaches used to do so span different methods, including probabilistic plan-based approaches (e.g., \cite{kabanza2010opponent}), Bayesian Networks (e.g., \cite{albrecht1998bayesian}), Hidden Markov Models (e.g., \cite{matsumoto2004mmog}), and Partially Observable Markov Decision Process (POMDPs) \cite{baker2014modeling}. While many approaches have been proposed, none of these approaches target behavior tendencies modeled through individual differences in a game, which is the goal of our current work. 

A great example that used a probabilistic plan-based approach to plan recognition is the work of Kabanza et al. \cite{kabanza2010opponent}. They developed a plan-based approach to probabilistically infer plans and goals in a strategy game. They used a Hierarchical Task Network to generate plans and then used these plans to infer what plans or goals players are taking given their behavioral data. While the approach showed some success, it results in low accuracy prediction rates and is intractable for most complex games. Bayesian Networks have been used in several games, especially adventure games~\cite{albrecht1998bayesian} to infer next action, and educational games~\cite{conati1997line,conati2002using} to infer knowledge or learning. However, none of these techniques produce high enough accuracy or have been tested in today's games. Further, they also do not model behavior tendencies which tend to vary across individuals and even across time.

The only work that investigated modeling individual variations to infer player types or personalities was the work of Bunian et al. \cite{bunian2017modeling}. They used HMMs to uncover individual differences between players using \textit{VPAL} (Virtual Personality Assessment Lab) game data. However, it does not focus on why players tend to exhibit such characteristics. The tendency of the players' observed behaviors to make decisions still remains elusive due to the lack of descriptive modeling of behaviors such as players' decision making.

Similar to our work is the work on using POMDPs to perform plan recognition~\cite{baker2014modeling}, belief modeling~\cite{baker2012bayesian}, and intent recognition~\cite{sukthankar2014plan}. Specifically, Baker and Tenanbaum's work ~\cite{baker2014modeling} is relevant as they develop a computational model to capture theory of mind using POMDPs. While such works have made progress towards enabling a computational approach to the theory of mind it does not explicitly model the behavior tendencies that contribute to a player's decision making process. Instead, decisions are viewed as a means to fulfil some desires or goals which are inverse inferred. Another sub-area of player modeling relevant to our approach is player decision modeling~\cite{holmgard2014generative,holmgaard2014evolving}. In these works, while the focus is on a player's decision-making behaviors, the purpose of modeling is generative~\cite{smith2011inclusive} implying that there is a lack of rationale for a given decision being made while the emphasis lies on the decisions being closely reproduced~\cite{bindewald2016clustering}. Such models has been applied to agents that act as play testers~\cite{liapis2015procedural}. However, there is lack of player decision modeling from a descriptive standpoint while maintaining the granularity and cognitive underpinnings of the modeled decisions. Thus, further work is required to model granular details of a player's decision making process towards understanding their behavioral tendencies.
\section{A Computational Cognitive Model of A Player's Decision Making}\label{sec:model}


\subsection{An Abstraction of A Player's Decision Making Process}

We make the following assumptions to abstract a player's decision making process. First, we assume that a player perfectly knows the game mechanism. Thus, there is no uncertainty stemming from a player's lack of knowledge about the game. This is equivalent to assuming that a player is an expert. Second, we assume that a player has bounded rationality and limited cognitive resources. This implies that a player does not think multiple steps ahead neither can they realize the end state of the map. Thus, a player's decision making is modeled myopically such that only the situational state of the game at any time step influences their subsequent decisions. Third, we assume a player can view the entire game map. This is equivalent to having a ``mini-map'' feature in a game that enables players to have a birds eye view of the map. Fourth, we do not model multiplayer interactions and consider individual player's decision making and cognitive behaviors.

We abstract a player's sequential decision making process as follows. We consider that a player makes two decisions $D = \left( X,Y \right)$ in a sequential manner. We assume that the decisions have discrete and finite outcomes such that the outcomes of decision $X$ influence the outcomes of decision $Y$. Such an abstraction enables us to consider sequential decision making in a parsimonious manner (minimum number of decisions required to create a sequence of decisions). We also assume to have the game state data $\mathcal{G}$ which holds relevant information about the state of the game when a player made the corresponding decisions. Thus, for a random number of samples $N$ of a player's decision making data across several sessions of their gameplay, we assume to have a set of player's sequential decisions $D_{1:N} = \left( X_{1:N},Y_{1:N} \right)$ and the corresponding game states $\mathcal{G}_{1:N}$ .


\subsection{Cognitive Modeling of A Player's Decision Making} \label{subsec:cogmod}



After abstracting a player's decision-making process, we model \textit{how} players make the specific decisions. To do so, we leverage decision theory to model decisions as functions of attributes or features of observed data within the game weighted differently by each individual player. Feature functions enable deterministic modeling that leverage game states to model decision attributes. Decision outcomes are modeled probabilistically using likelihood functions, with function parameters such as an individual's behavior tendency $\boldsymbol{\theta}$, which adds stochasticity in the predictions. Our modeling approach acknowledges that players make decisions subjectively based on their behavior tendency or individual preferences. Moreover, the assumption of probabilistic decisions assumes the limited cognitive ability of a player to make accurate decisions even though their judgments may be aligned with rational judgments.

Formally, we refer to a mapping between the observed game data to some situational factor as a \textit{feature function}. A feature function (or simply feature) incorporates the observed game state $\mathcal{G}_n$ for a data sample $n$ into the decision models. Given that multiple situational factors may influence decisions, a decision strategy is specified in terms of a weighted sum of multiple independent features. Moreover, a threshold value is associated with each feature to model an individual's mental activation to the subjective strength of a particular feature. 

Mathematically, we characterize a decision strategy for a decision $X_{n}$ in a data sample $n$ with $O$ discrete outcomes $\{x_{n,1},\ldots,x_{n,O}\}$ using $R$ independent attributes or features denoted by $g_{1,x}(\mathcal{G}_n), \ldots , g_{R,x}(\mathcal{G}_n)$. The values of the features can be dependent on the decision alternative in consideration. Then, we model the stochastic decision process as follows:

\begin{equation}\label{eqn:generalX}
  \resizebox{\columnwidth}{!}{$   X_{n} = \begin{cases}
        x_{n,o},&\text{with probability}\;\operatorname{softmax}_{x_o}\left(\sum_{r=1}^{R}w_{r}\left(g_{r,x}(\mathcal{G}_n)-\delta_{r}\right)\right)
    \end{cases}$}
\end{equation}

and, the outcome probability is given by,
\begin{equation}\label{eqn:generalXprob}
 \resizebox{\columnwidth}{!}{$ \operatorname{softmax}_{x_o}\left(\sum_{r=1}^{R}w_{r}\left(g_{r,x}(\mathcal{G}_n)-\delta_{r}\right)\right)=\dfrac{\exp \left(\sum_{r=1}^{R}w_{r}\left(g_{r,x_o}(\mathcal{G}_n)-\delta_{r}\right)\right)}{\sum\limits_{\forall X} \exp \left(\sum_{r=1}^{R}w_{r}\left(g_{r,x}(\mathcal{G}_n)-\delta_{r}\right)\right)}. $}
\end{equation} where, $X_{n}=x_{n,o}$ is the observation that an individual chose an alternative $x_{n,o}$ for a decision $X$ for a data sample $n$, $\boldsymbol{\theta}= \{w_{1:R},\delta_{1:R}\}$ are player-specific cognitive parameters modeled as the feature weight and feature threshold parameters. The weight parameter $w_{r}$ can be positive or negative depending on whether an increase in $g_{r,x}(\mathcal{G}_n)$, respectively, increases or decreases the probability of an outcome. The threshold parameters imply that the weighted sum of the situational factors \textit{as activated by a player} determines their decision strategy. 

The decision strategy for a decision $Y_n$ can also be defined using a similar approach as discussed for the decision $X_n$ in Equation~\ref{eqn:generalX}. The only difference in consideration of the outcome probabilities of $Y_n$ will be that they are conditioned on the outcomes of $X_n$.


\subsection{Inverse Inference} \label{subsec:inverse}
Here, we mention a general strategy for inverse Bayesian inference. However, we note that it's important to consider game-specific causal models that represent the influence of situational factors on a player's decision making. Given decision data $D_{1:N}$, a prior over $\theta$, $p(\theta)$, a prior over game states $\mathcal{G}_{1:N}$, and the outcome probabilities for decision $X_n$, $p(X_n|\theta, \mathcal{G}_{1:N})$ and the outcome probabilities for decision $Y_n$, $p(Y_n|\theta, X_n, \mathcal{G}_{1:N})$, inference over posterior of $\theta$ is given by Bayes' rule:

\begin{equation}
 \resizebox{\columnwidth}{!}{$  p(\boldsymbol{\theta} | D_{1:N},\mathcal{G}_{1:N}) \propto p(X_{1:N}|\boldsymbol{\theta},\mathcal{G}_{1:N})p(Y_{1:N}|\boldsymbol{\theta},X_{1:N}, \mathcal{G}_{1:N}) p(\boldsymbol{\theta})p(\mathcal{G}_{1:N}).$}
\end{equation} We illustrate this inference in further detail in our use case in Section~\ref{subsec:inv_inf}.

\section{Case Study: \textit{BoomTown} Game} \label{sec:boomtown}

\begin{figure}[hbt!]
\includegraphics[width=8.5cm, height=5cm]{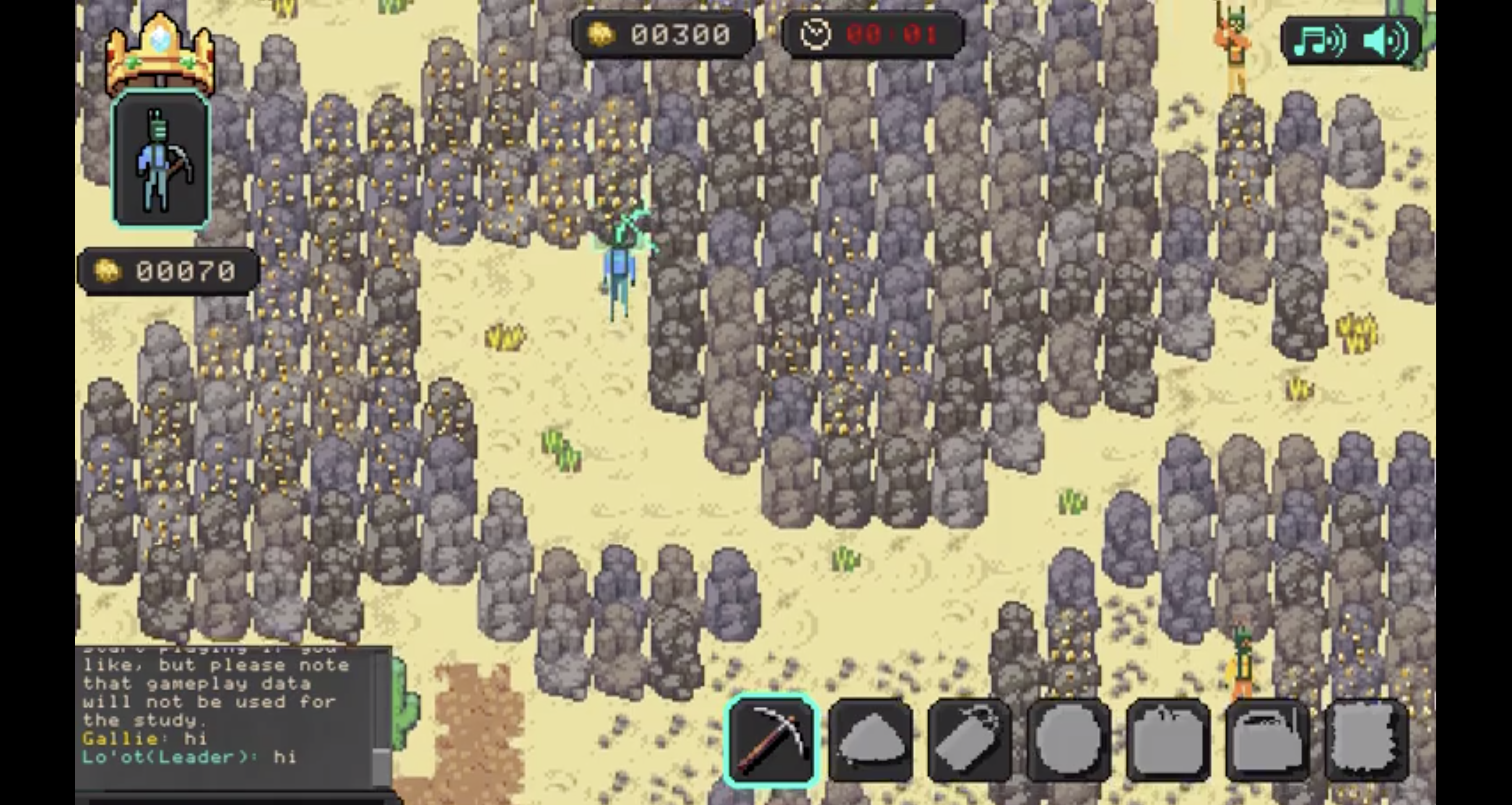}
\caption{Screenshot of \textit{Boomtown}.}
\label{fig: boomtown}
\centering
\end{figure}

We implemented our model in a game called \textit{BoomTown} which is a resource acquisition game developed by Gallup. Figure \ref{fig: boomtown} is a screenshot of the game. The objective of \textit{BoomTown} is for players to maximize their collection of ``gold nugget'', a game-specific resource. Players can do so by exploring a game map. The map is constructed through fundamental units called tiles. There are four different tiles including road tile, rock tile, gold nugget tile and obstacle tile. An agent can be physically present only on a road tile. The rock tile and the gold tile can be destroyed by the player while the obstacle tile is immovable and cannot be destroyed. Players can destroy the rock tile and the gold tile through the use of several items. By destroying gold tiles, players get a certain amount of gold per gold tile. A destroyed rock or gold tile becomes road tile. As players collect gold nuggets, a counter updates the amount of gold overall collected by the team. The game ends after a fixed period of time. 

We chose \textit{BoomTown} because, (1) the game is a single objective resource collection game which enables us to abstract player's decision making process as a simple sequential decision making scenario. This enables us to focus on the cognitive modeling aspect of computational player modeling as opposed to modeling the complexity of a game environment, (2) we have human data on gameplay behaviors which can be leveraged in the long run to run validation studies on the proposed computational model in this study, and (3) the game has multi-player mode which also provides flexibility to build our model in future work for more complex scenarios that match esports-like contexts.

\subsection{Modeling \textit{BoomTown} Decisions}\label{subsec:BtDecisions}

For \textit{BoomTown}, we consider players' sequential moving decisions. These include the decision to move or not $M_n$ and which direction to move $D_n$. At any point in the game, a player decides whether they want to move in the map or not. If they decide to move, then they need to decide which direction to move. If they choose to stay, they decide to use items in the game to mine gold. We do not model the use items decision. We only model the moving decisions to illustrate the sequential nature of the decision making process.

We note that the first decision to move or not has two outcomes. We consider two significant situational features in the game for a player's decision to move: the rock tile and the gold nugget tile. Thus, $R=2$ for our move decision model. For other games, model developers would need to investigate the situational factors that influence players' gameplay. The first situational factor is \textit{Gold Around}, $GA_n=g_{1}(\mathcal{G}_n)$ which represents how many gold nuggets are around a player within some map region for a data sample $n$. The second situational factor is \textit{Rock Around}, $RA_n =g_{2}(\mathcal{G}_n)$ which represents how many rocks are around a player within some map region for a data sample $n$. The calculation of these features is independent of the alternative to move or not thus we drop the term $x$ in $g_{r,x}(.)$ as discussed in Section~\ref{subsec:cogmod}. We model the stochastic moving process $M_n$ for \textit{BoomTown} as follows:

\begin{equation}\label{eqn:boommove}
 \resizebox{\columnwidth}{!}{$   M_{n} = \begin{cases}
        1,&\text{with probability}\;\operatorname{sigm}\left(\sum_{r=1}^{2}w_{r}\left(g_{r}(\mathcal{G}_n)-\delta_{r}\right)\right)\\
        0,&\text{otherwise},
    \end{cases}$}
\end{equation}
and, the moving probability is given by,
\begin{equation}
\resizebox{\columnwidth}{!}{$    p(M_n=1|\mathcal{G}_n)= \dfrac{1}{1+\exp{\left(w_1(GA_n-\delta_{1})-w_2(RA_n-\delta_{2})\right)}}$},
\end{equation} where, without loss of generality, the weight parameter $w_{1:2}$ are considered to be positive while the negative or positive influence of each of the features on the moving probability is intuitively coded. Thus, increase in gold around the player decreases the probability to move while increase in rock around the player increases their probability to move. Moreover, the sigmoid function $\operatorname{sigm}()$ is a special case of the $\operatorname{softmax}()$ function discussed in Equation~\ref{eqn:generalX} for a decision with two outcomes.



The second decision of where to move $D_n$ is considered to have five alternatives, namely, North ($d_{n,1}$), South ($d_{n,2}$), East ($d_{n,3}$), West ($d_{n,4}$), and no direction ($d_{n,5}$) . Thus, $D_n=\{d_{n,1:5}\}$. Moreover, we consider five situational factors such that $R=5$. The situational factors are dependent on a direction alternative $d_{n,i}$ and include \textit{Gold Around}, $GA_{n,i}=g_{1,i}(\mathcal{G}_n)$, and \textit{Rock Around}, $RA_{n,i} =g_{2,i}(\mathcal{G}_n)$ which represents how many gold nuggets and rocks are around a player in a direction $d_{n,i}$ for a data sample $n$. There are three additional situational factors, namely, $GD_{n,i}=g_{3,i}(\mathcal{G}_n)$ the average distance of the \textit{gold around}, $RD_{n,i}=g_{4,i}(\mathcal{S}_t)$ the average distance of the \textit{rock around} the player, and $OA_{n,i}=g_{5,i}(\mathcal{G}_n)$ the obstacle tiles around player position in direction $d_{n,i}$. 

\begin{figure}[htb!]

\includegraphics[width=0.48\textwidth]{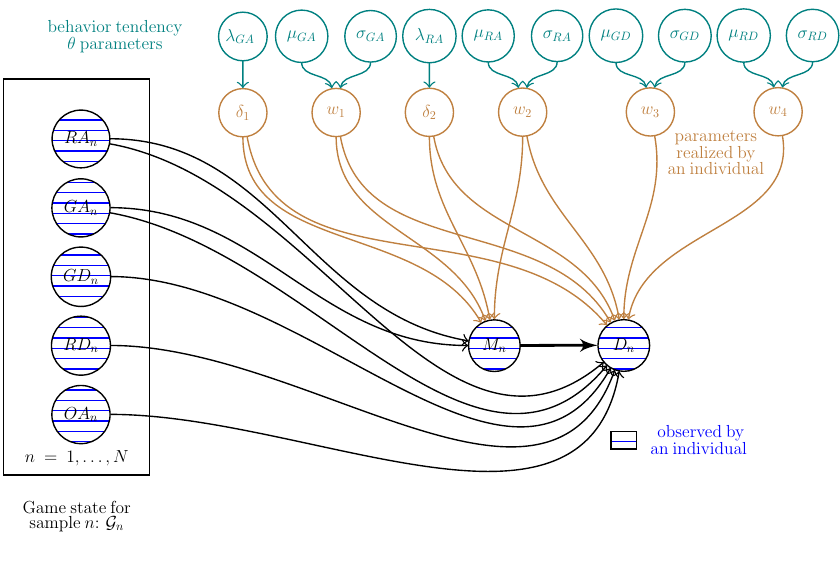}

\caption{Graphical illustration of the decision to move $M_{n}$ and where to move $D_{n}$ influenced by the game data $\mathcal{G}_n$ and individual specific parameters $\boldsymbol{\theta}$. The parameters $w_{1:4}$ and $\delta_{1:2}$ are cognitively realized by the individual. The parameters $\lambda_{GA}$, $\mu_{GA}$, $\sigma_{GA}$, $\lambda_{RA}$, $\mu_{RA}$, $\sigma_{RA}$, $\mu_{GD}$, $\sigma_{GD}$, $\mu_{RD}$, and $\sigma_{RD}$ are a part of an individual's behavior tendency $\boldsymbol{\theta}$. }
\label{fig:abstraction}
\end{figure}

The player moves in no direction $D_n = d_{n,5}$ given the decision to move $M_n=0$. Thus, the stochastic decision of which direction $D_n$ to move, given the decision to move $M_n=1$, is modeled as follows:


\begin{equation}\label{eqn:where}
 \resizebox{1.0\hsize}{!}{ $  D_{n} = \begin{cases}
       d_{n,1},&\text{with probability}\;\operatorname{softmax}_{d_1}\left(w_{1:R},g_{r,d}(\mathcal{G}_n)\right)\\
         d_{n,2},&\text{with probability}\;\operatorname{softmax}_{d_2}\left(w_{1:R},g_{r,d}(\mathcal{G}_n)\right)\\
         d_{n,3},&\text{with probability}\;\operatorname{softmax}_{d_3}\left(w_{1:R},g_{r,d}(\mathcal{G}_n)\right)\\
         d_{n,4},&\text{with probability}\;\operatorname{softmax}_{d_4}\left(w_{1:R},g_{r,d}(\mathcal{G}_n)\right)\\
        
    \end{cases}$}
\end{equation} where, the probability to move in a direction is given by,

\begin{equation}
   \resizebox{1\hsize}{!}{ $   p(D_n=d|M_n=1,\mathcal{G}_n,\boldsymbol{\theta}) =  \operatorname{softmax}_{d_{i}}\left(w_{1:R},g_{r,d}(\mathcal{G}_n)\right), 
   $}
\end{equation} such that,

\begin{equation}\label{eqn:boom_direc}
  \resizebox{1\hsize}{!}{ $      \operatorname{softmax}_{d_{i}}\left(.\right) = \dfrac{ \mathbbm{1}_{0} \left(  g_{5,d}(\mathcal{G}_n) w_5\right)\exp \left(\sum_{r=1}^{2} \dfrac{w_r}{w_{r+2}}\dfrac{g_{r,d_i}(\mathcal{G}_n)}{g_{r+2,d_i}(\mathcal{G}_n)}\right)}{\sum\limits_{d=1}^{d=4} \mathbbm{1}_{0} \left(  g_{5,d}(\mathcal{G}_n)w_5\right)\exp \left(\sum_{r=1}^{2} \dfrac{w_r}{w_{r+2}}\dfrac{g_{r,d}(\mathcal{G}_n)}{g_{r+2,d}(\mathcal{G}_n)}\right)}. 
  $}
\end{equation}

where, $D_n=d$ is the observation that an individual moves in direction $d$ on the map for the $n^\text{th}$ sample, $\boldsymbol{\theta}= \{w_{1:R}\}$ are player-specific cognitive parameters modeled as feature weights for $R=5$. The weight parameter $w_{r}$ can be positive or negative depending on whether an increase in $g_{r,i}(\mathcal{G}_n)$, respectively, increases or decreases the probability of moving. We consider the average distance of the gold and rock in a particular direction to be inversely proportional to an individual's utility to move in that direction. Moreover, the ratio of the total gold amount to the average gold distance in a particular direction signifies that an individual player would balance exploration of large gold clusters further away with the exploitation of smaller clusters closer to them. Similarly for rock, a player would be averse to large rock clusters in vicinity and prefer less rock dense areas. The feature $g_{5,d}(\mathcal{G}_n)$ indicates if there is an obstacle in the cell next to a player in direction $d$. Thus, the indicator function $\mathbbm{1}_{0}\left( g_{5,d}(\mathcal{G}_n)w_5\right)$ only avails the directions where a player can move. Consequently, we drop $w_5$ as obstacles are a part of game mechanics that affects each player in the same way. Moreover, we set the threshold parameters $\delta_r=0$ to model where to move. This is because the player has already decided to move and the minimum threshold required to move in any direction is a positive value of the features. Also, it's the \textit{relative} evaluation of the gold and rock amount to their average distances which is of importance to a player's direction decision. We note that we deliberately wrote Equation~\ref{eqn:boom_direc} in a different form than Equation~\ref{eqn:generalXprob} to highlight that game-specific considerations will influence how model developers choose to represent decision models towards better explainability. Figure~\ref{fig:abstraction} illustrates the behavior tendency parameters $\boldsymbol{\theta}$ as utilized in \textit{BoomTown}.

\subsection{Inferring A Player's Cognitive Variables from Data }\label{subsec:inv_inf}

In this section, we discuss how to infer an individual's modeled cognitive parameters, that is, their $\boldsymbol{\theta}$ given $N$ samples of the game state $\textbf{g}_{1:N}$ and the decision data history $h_{1:N}=\{m_{1:N},d_{1:N}\}$ which includes the player's data for their decisions to move $m_n$ and where to move $d_n$ for each sample $n$.

We proceeded in a Bayesian way which required the specification of a prior $p(\boldsymbol{\theta})$ for $\boldsymbol{\theta}$, a prior $p(\textbf{g}_{1:N})$ for game state, a likelihood $p(h_{1:N}|\boldsymbol{\theta})$ for decisions to move $m_{1:N}$ and where to move $d_{1:N}$ given $\boldsymbol{\theta}$. The posterior state of knowledge about $\boldsymbol{\theta}$ is simply given by Bayes' rule:
\begin{equation}\label{eqn:BM_posterior}
    p(\boldsymbol{\theta} | h_{1:N},\textbf{g}_{1:N}) \propto p(h_{1:N}|\boldsymbol{\theta},\textbf{g}_{1:N}) p(\boldsymbol{\theta})p(\textbf{g}_{1:N}),
\end{equation}
and we characterized it approximately via sampling.
We now describe each of these steps in detail.

We associate behavior tendency with the vector of parameters $\boldsymbol{\theta} = \{ w_{1:R}, \delta_{1:R}\}$ defined in Section \ref{subsec:BtDecisions}. We describe our prior state of knowledge about $\boldsymbol{\theta}$ by assigning a probability density function such that it becomes a random vector modeling our epistemic uncertainty about the actual cognition of the individual. Having no reason to believe otherwise, we assume that all components of an individual's behavior tendency are a priori independent, i.e., the prior probability density (PDF) factorizes as:

\begin{equation}
    p(\boldsymbol{\theta}) = \prod_{r=1} ^{r=4} p(w_r)\prod_{r=1} ^{r=2}p(\delta_r),
\end{equation} where, $p(w_r)$ is assigned an uninformative Jeffrey's prior, i.e., $p(w_r) \propto \frac{1}{w_r}$, and 
\begin{equation}\label{eqn:priorb}
\begin{array}{ccc}
    \delta_{1}&\sim& \mathcal{N}(50,25),\\
    \delta_{2}&\sim& \mathcal{N}(50,25).\\
    \end{array}
\end{equation} The mean and standard deviation of the normal distribution for the threshold priors were chosen based on the game ranges for rock and gold values.

The game state $\textbf{g}_{1:N}$ is a vector of features sampled randomly for a sample $N$ and is thus assumed to have a uniform distribution. Such an assumption enables us to circumvent the problem of modeling the game mechanics where player actions influence the game states. Thus, we note that $N$ samples of game data are equivalent to sampling decision data $m_{1:N}$ and $d_{1:N}$ of a player from $N$ randomly generated map scenarios. In general, derivation of game specific prior probabilities for game states will require understanding of the game mechanics that govern the initialization of game states for a game map.

The likelihood $p(h_{1:N}|\boldsymbol{\theta},\textbf{g}_{1:N})$ is calculated conditioned on $\boldsymbol{\theta}$ and $\textbf{g}_{1:N}$. We have:

\begin{equation}
    p(h_{1:N} | \boldsymbol{\theta},\textbf{g}_{1:N}) = \prod_{q=1}^N p(h_q | \boldsymbol{\theta},\textbf{g}_q),
\end{equation} given the independent sampling assumption of our model For each product term, we have:

\begin{equation}\label{eqn:likelihood}
    p(h_q|\boldsymbol{\theta} , \textbf{g}_q) = p(m_q|\textbf{g}_q,\boldsymbol{\theta}) p(d_q|m_q,\textbf{g}_q,\boldsymbol{\theta}).
\end{equation}

The first term in Equation~\ref{eqn:likelihood} is:

\begin{equation}
  \resizebox{\columnwidth}{!}{$
  \begin{array}{cc}
    p(m_q| \textbf{g}_{q},\theta)=   &  \left[\operatorname{sigm}\left(\sum_{r=1}^{R}w_{r}\left(g_{r}(\textbf{g}_{q})-\delta_{r}\right)\right)\right]^{m_q} \\
       &  \left[1 -\operatorname{sigm}\left(\sum_{r=1}^{R}w_{r}\left(g_{r}(\textbf{g}_{q})-\delta_{r}\right)\right)\right]^{1-m_q},
  \end{array} 
$}
\end{equation} where, weights $w$ and threshold $\delta$ parameters are conditioned on $\theta$. This equation is derived from Equation~\ref{eqn:boommove}.

The second term is:

\begin{equation}
    \resizebox{\columnwidth}{!}{$ 
  \begin{array}{cc}
      p(d_q|m_q,\textbf{g}_{q},\theta)=   &  \left[\operatorname{softmax}_{d_1}(\theta,\textbf{g}_{q,r,d})^{d_{1,q}} \operatorname{softmax}_{d_2}(\theta,\textbf{g}_{q,r,d})^{d_{2,q}} \right] ^{m_q} \\
         & \left[\operatorname{softmax}_{d_3}(\theta,\textbf{g}_{q,r,d})^{d_{3,q}}\operatorname{softmax}_{d_4}(\theta,\textbf{g}_{q,r,d})^{d_{4,q}}\right] ^{m_q}
    \end{array},$}
\end{equation} where, $d_q=\{d_{1,q},d_{2,q},d_{3,q},d_{4,q}\}$ is $1$ or $0$ when a player moves in one of the directions or not for each sample $q$. We note that when a player chooses to stay then $m_q=0$ and the decision of where to move is not relevant. This equation is derived from Equation~\ref{eqn:where}.

\section{ Verification Strategy and Results} \label{sec:results}

\subsection{Synthetic Data Generation}

We generate game play data by simulating the model discussed in Section~\ref{subsec:BtDecisions} with $N=5000$. We considered two behavior tendencies, (1) rock agnostic tendency, and (2) a rock aversion tendency. A player with rock agnostic tendency is one who attributes a lot of consideration to large gold clusters and is agnostic about the amount of rock structures. On the other hand, a player with rock averse tendency focuses more on the gold in the rock-free regions such that they would not have to mine through the rocks to reach to the gold.

Our model enables capturing such player tendencies through an initialization of the weight and threshold parameters. Table \ref{table:Player Cognition Settings} tabulates the initialized parameters for the two behavior tendencies. We note that the differentiating parameter for the two tendencies is $w_2$. This parameter quantifies the tendency of a player to consider the rock around. We note that a rock agnostic tendency doesn't focus much on the rock around $w_2=0.3$ whereas the rock averse tendency attributes high weight to the rock around $w_2=0.8$ implying an avoidance to rock clusters. We note that since the game objective is to collect gold, both behavioral tendencies have a high weight $w_1$ on gold. The threshold parameters $\delta_{1,2}$ focus on a player's emphasis on the size of the rock and gold clusters but they do not explain the tendencies of interest.





\begin{table}[tbh!]
\caption{Player behavior tendency settings.}
\label{table:Player Cognition Settings}
\centering

\resizebox{\columnwidth}{!}{
\begin{tabular}{|c|c|a|c|c|c|c|} \hline
\diagbox[font=\footnotesize,innerwidth=2.1cm]{\textbf{Tendency}}{$\boldsymbol\theta$} & \textbf{$w_{1}$} & \textbf{$w_{2}$} & \textbf{$\delta_{1}$} & \textbf{$\delta_{2}$} & \textbf{$w_{3}$} & \textbf{$w_{4}$}\\ \hline

Rock Agnostic Tendency & 0.90 & 0.30 & 20.00 & 60.00 & 1.13 & 1.00 \\ \hline

Rock Averse Tendency & 0.95 & 0.80 & 50.00 & 20.00 & 3.17 & 1.14 \\ \hline 
\end{tabular}}
\end{table}

\subsection{Inverse Inference Using MCMC}

We sampled from the posterior (Equation~\ref{eqn:BM_posterior}) using the No-U-Turn Sampler (NUTS)~\cite{hoffman2014no}, a self-tuning variant of Hamiltonian Monte Carlo~\cite{duane1987hybrid} from the PyMC3~\cite{salvatier2016probabilistic} Python module. We ran two chains of the (Markov Chain Monte Carlo) MCMC simulations and for each chain we ran $10000$ iterations with a \textit{burn-in} period of $2000$ samples that are discarded.

\subsection{Results}\label{subsec:results}
\begin{figure}[hbt!]
\includegraphics[width=8.5cm, height=6.5cm]{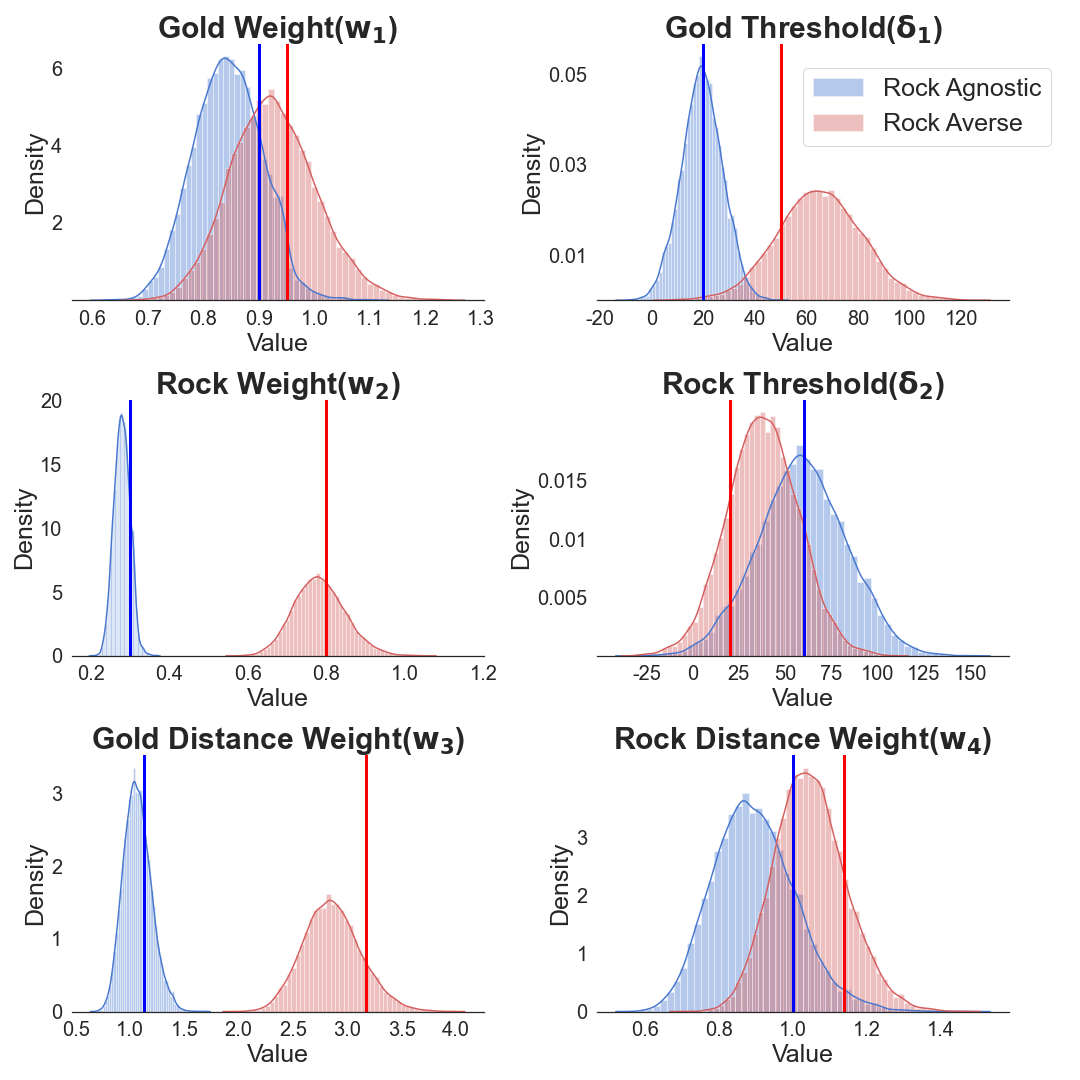}
\caption{Posterior of parameters for rock agnostic and rock averse tendencies.}
\label{fig: Gold Greedy Player Posterior Distributions of Theta}
\centering
\end{figure}

We infer the behavioral tendency parameters for both the simulated datasets and are able to differentiate the behavioral data using the inferred parameters. Specifically, the posterior of $w_2$ differentiates the two tendencies of interest as intended from the simulated data sets. Figure \ref{fig: Gold Greedy Player Posterior Distributions of Theta} shows the posteriors distributions over each modeled parameter. The blue and red vertical lines represents the setting of the rock agnostic and rock averse tendency used for data simulation, respectively. 

\begin{table}[tbh!]
\caption{Rock agnostic tendency summary statistic values.}
\label{table:Gold Greedy Player Statistic Values}
\centering
\tiny
\resizebox{\columnwidth}{!}{
\begin{tabular}{|c|c|c|c|c|} \hline
\diagbox[font=\tiny,innerwidth=2.1cm]{\textbf{Variables}}{\textbf{Statistics}} & \textbf{$mean$} & \textbf{$sd$} & \textbf{$hdi3\%$} & \textbf{$hdi97\%$}\\ \hline

$w_{1}$ & 0.84 & 0.06 & 0.73 & 0.95 \\ \hline
\rowcolor{LightCyan}
$w_{2}$ & 0.28 & 0.02 & 0.24 & 0.31 \\ \hline 

$\delta_{1}$ & 19.52 & 7.91 & 4.17 & 33.88 \\ \hline

$\delta_{2}$ & 59.19 & 23.88 & 13.44 & 103.02 \\ \hline 

$w_{3}$ & 1.13 & 0.09 & 0.71 & 1.06 \\ \hline

$w_{4}$ & 0.97 & 0.03 & 0.26 & 0.37 \\ \hline 
\end{tabular}}
\end{table}

\begin{table}[tbh!]
\caption{Rock averse tendency summary statistic values.}
\label{table:Rock Reject Player Statistic Values}
\centering
\tiny
\resizebox{\columnwidth}{!}{
\begin{tabular}{|c|c|c|c|c|} \hline
\diagbox[font=\tiny,innerwidth=2.1cm]{\textbf{Variables}}{\textbf{Statistics}} & \textbf{$mean$} & \textbf{$sd$} & \textbf{$hdi3\%$} & \textbf{$hdi97\%$}\\ \hline

$w_{1}$ & 0.93 & 0.08 & 0.79 & 1.08 \\ \hline
\rowcolor{LightCyan}
$w_{2}$ & 0.78 & 0.06 & 0.66 & 0.90 \\ \hline 

$\delta_{1}$ & 64.93 & 16.25 & 34.05 & 94.99 \\ \hline

$\delta_{2}$ & 37.86 & 19.22 & 1.21	& 73.31 \\ \hline 

$w_{3}$ & 2.97 & 0.01 & 0.30 & 0.35 \\ \hline

$w_{4}$ & 1.06 & 0.03 & 0.69 & 0.80 \\ \hline 
\end{tabular}}
\end{table}

\begin{figure}[hbt!]
\includegraphics[width=8.5cm, trim=0cm 0cm 0 2cm , clip]{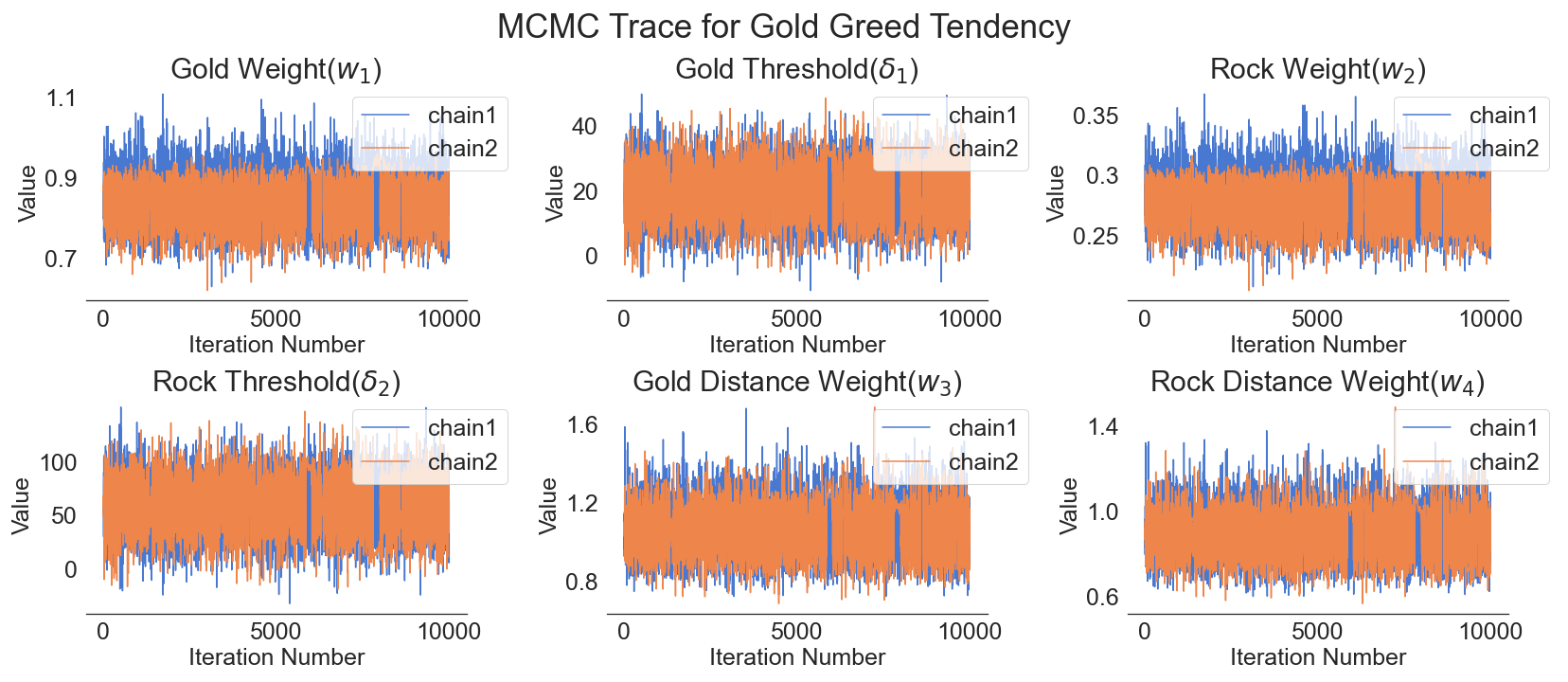}
\caption{MCMC traces of two chains for rock agnostic tendency.}
\label{fig: Gold Greedy Player MCMC Trace}
\centering
\end{figure}


\begin{figure}[hbt!]
\includegraphics[width=8.5cm, trim=0cm 0cm 0 1.1cm , clip]{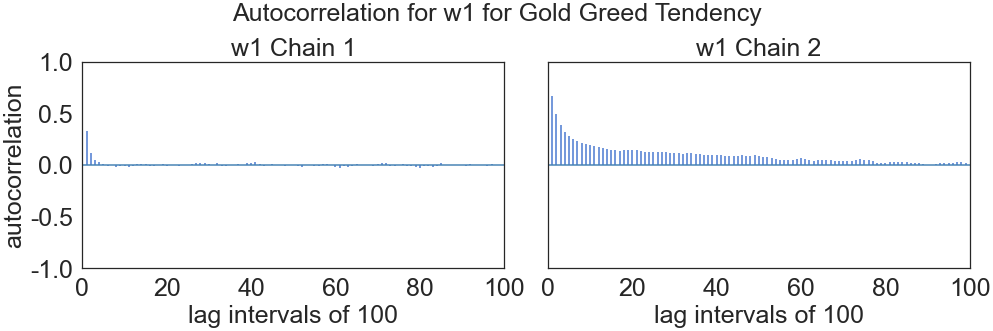}
\caption{Autocorrelation of $w_1$ for rock agnostic tendency.}
\label{fig: Posterior Autocorrelation for Gold Greedy Player}
\centering
\end{figure}

For rock agnostic tendency, Table \ref{table:Gold Greedy Player Statistic Values} shows the statistical summary data from MCMC simulation.The (highest density interval) hdi\%3 and hdi97\% show the range of points of distribution which is credible. We find that the weight estimations have less standard deviation than the threshold estimations. The narrow range between hdi3\% and hdi97\% also represents the certainty of belief on weight estimations. Figure \ref{fig: Gold Greedy Player MCMC Trace} shows the sampling process. Figure \ref{fig: Posterior Autocorrelation for Gold Greedy Player} shows a representative autocorrelation of all six cognitive variables. The low autocorrelation at the end shows the convergence of $w_{1}$. The same is also true for other cognitive variables. We find similar results for rock averse players and we only show the summary statistics in Table \ref{table:Rock Reject Player Statistic Values}.


\section{Conclusions and Future Work}\label{sec:discussion}

The proposed model in this study serves as a stepping stone towards inferring player cognition in digital games. Currently, our model is only verified to retrieve cognitive parameters from synthetically generated data which does not have any noise such as deviations from the modeled decision making strategy. Thus, future work includes testing our model with human subjects data to validate the generalizability of the modeled cognitive parameters across several games where moving decisions are made by players. 

Our model does not account for several other decisions that players make such as drafting a team or selecting the resources or items used within a gameplay. However, through this study we provide a foundation towards modeling such decisions and extending our model to account for other decisions and decision making processes. For example, in this study, we assume two sequential decisions. This assumption can be relaxed by considering greater number of decision sequences where each decision is conditionally dependent on previous decisions. This would increase the number of nodes illustrated in Figure~\ref{fig:abstraction}. However, the specifics of the decisions and the sequences will be dependent on the game mechanics, the level of abstraction of player behaviors, and the game-specific processes. 

Our model also does not consider multiple players which would be crucial in several theory of mind contexts. Currently, our model is assumed to be a spectator for a single player who may engage in practice sessions and receive feedback about the model's theory of mind for their gameplay. Moreover, further work is required to transform the inferences about player cognition to explainable rationales which would require further investigations on rationale generation in context of the theory of mind.


\bibliographystyle{aaai}
\bibliography{biblio.bib}

\end{document}